\documentclass[aps,pre,twocolumn,showpacs,showkeys,groupedaddress]{revtex4}
\usepackage{graphicx} 
\usepackage{rotating}
\usepackage{amssymb}      
\usepackage{amsmath}     
\usepackage{epsfig}
\usepackage[normalem]{ulem}     
\usepackage{bm}     
\usepackage{color}

\begin{document}

\title{Replicate Periodic Windows in the Parameter Space of Driven Oscillators}

\def\active{0}

\author{E. S. Medeiros$^1$, S. L. T. de Souza$^2$, R. O. Medrano-T.$^{1,3}$, I. L. Caldas$^1$} 

\affiliation{$^1$ Instituto de F\'isica, Universidade de S\~ao Paulo, S\~ao Paulo, Brasil}

\affiliation{$^2$ Universidade Federal de S\~ao Jo\~ao del-Rei, Campus Alto Paraopeba, Minas Gerais, Brazil}

\affiliation{$^3$ Departamento  de  Ci\^encias   Exatas  e  da  Terra,
Universidade  Federal de S\~ao  Paulo, Diadema, S\~ao Paulo, Brasil}

\begin{abstract}

In the bi-dimensional parameter space of driven oscillators, shrimp-shaped periodic windows are immersed in chaotic regions. For two of these oscillators, namely, Duffing and Josephson junction, we show that a weak harmonic perturbation replicates these periodic windows giving rise to parameter regions correspondent to periodic orbits. The new windows are composed of parameters whose periodic orbits have periodicity and pattern similar to stable and unstable periodic orbits already existent for the unperturbed oscillator. These features indicate that the reported replicate periodic windows are associated with chaos control of the considered oscillators.

\vglue 0.4 
truecm

\end{abstract}


\maketitle

\section{Introduction}
\label{sec:intro}

The parameters of deterministic dynamical systems play an important role to specify the transitions between chaotic and periodic behavior. This parameter influence on the attractor transition can be represented in the parameter space \cite{1,2,3,4,5}. In particular, for both discrete-time \cite{1,6} and continuous dynamical systems \cite{7}, it is known that the set of parameters for which a system exhibit periodic behavior are periodic windows immersed in chaotic regions, in the form of shrimps, in the bi-dimensional parameter space. In the past two decades, periodic windows have been numerically obtained for a large number of applied dynamical systems like lasers \cite{8,9}, electronic circuits \cite{9,10,11}, mechanical oscillators \cite{12}, and also in population dynamics \cite{13}. Periodic windows have also been identified in experiments with electronic circuits \cite{14, 15, 16}.

Additionally, the bi-dimensional parameter diagram is an important tool to analyze the outcome of techniques used to control chaos for allowing a global overview of periodic and chaotic behavior \cite{17}. In literature, different approaches have been proposed to control chaos. Essentially, two main methods have been successfully  employed in several dynamical systems, namely the feedback and the nonfeedback methods. The feedback methods are applied to maintain the trajectory in a desired unstable periodic orbit embedded in the chaotic attractor \cite{18}. In contrast, the nonfeedback methods eliminate the chaotic behavior by a slightly modification in the system dynamics \cite{19}. The suppression of chaos by applying an external weak perturbation is an interesting application of a nonfeedback method. 

Suppression of chaos by adding a weak harmonic perturbation has been numerically and theoretically reported specially for systems with an original harmonical driven \cite{20}. For example, a weak harmonic perturbation has been used to suppress chaos in the forced Duffing oscillator \cite{21}, and a similar perturbation has been applied to control chaos in a Josephson junction oscillator \cite{22}. Considering that in \cite{20,21,22} the controlling of these two well-known systems were essentially accomplished for limited ranges of system parameters, a further parameter space analysis considering a weak harmonic perturbation is required. 

The main motivation of this article is to determine the alterations in the parameter space due to weak harmonic perturbations applied to continuous-time dynamical systems. To investigate that we analyze how such perturbations modify the attractors and the parameter space of the harmonically driven Duffing and Josephson oscillators.  For the first time in the literature, we observe that this harmonic perturbation replicates the periodic windows in the parameter space of these systems. Moreover, we find evidences that new periodic orbits, whose parameters are in the new periodic windows, are similar to unstable periodic orbits embedded in the unperturbed chaotic attractor and periodic orbits already existing. We discover that the reported replicate periodic windows are associated with chaos control of the considered oscillators.

This letter is organized as follows: In Section~\ref{sec:duff}, we obtain the parameter space of the unperturbed Duffing oscillator and we establish the suitable parameters to implement the weak perturbation. In Section~\ref{sec:perduff}, for the Duffing oscillator, we investigate the periodicity, shape and Lyapunov exponents of the perturbed periodic and of the unperturbed chaotic orbit. In Section~\ref{sec:pertjose}, we present, for the Josephson oscillator, another example of replicate periodic windows caused by a weak periodic perturbation. Finally, in Section 5 we summarize our main conclusions.

\section{Parameter space structure of the Duffing Oscillator}
\label{sec:duff}
The Duffing oscillator is a well-known model to describe oscillations of a mass obeying a fourth order symmetric potential \cite{23}. This system has a large number of applications, specially to model physical systems, and has been extensively studied in theoretical, numerical and experimental approaches \cite{24,25,26}. Here, we consider a simple version of the Duffing system, which describe oscillations of a single-well potential. 

The time evolution of this system is determined by solution of the following dimensionless equation:

\begin{equation}
\ddot{x}+c \dot{x}+x^3=\beta \cos (\omega t).
\label{eq:1}
\end{equation}

Here, the parameter $c$ is the amplitude of the system damping, $\beta$ is the forcing amplitude, and $\omega$ is the natural system frequency settled at $\omega=1.0$. We numerically obtain solutions of the Duffing equation by using a fourth-order Runge-Kutta method with fixed step $h=0.001$. 

Additionally, to investigate transitions between chaotic and periodic attractors of the Duffing oscillator, the largest non zero Lyapunov exponent is obtained \cite{27} for each point of a bi-dimensional grid of the system parameters (parameter space).

\begin{figure*}[!htp]
\centering
\includegraphics[width=17cm,height=9cm]{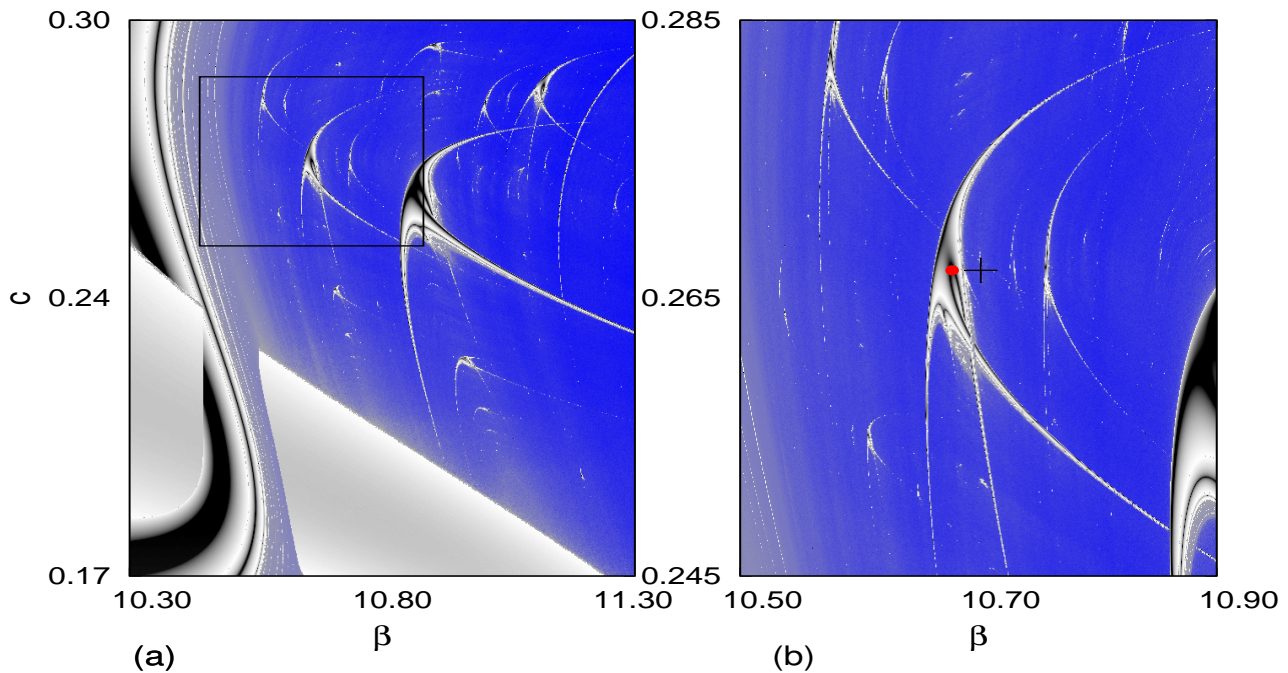}
\caption{ (Color online) (a) Bi-dimensional ($c$ $\times$ $\beta$) parameter space of the unperturbed Duffing oscillator, for $\omega = 1.0$. (b)  Magnification of the squared area. The red circle indicates periodic behavior at $\beta=10.6780$ and $c=0.2670$, the black plus symbol indicates chaotic behavior at $\beta = 10.7020$ and $c = 0.2670$.}
\label{fig:1}
\end{figure*}

In Figure~\ref{fig:1}, we show the bi-dimensional parameter space ($c$ $\times$ $\beta$) obtained for the Duffing oscillator (also shown in \cite{25}). The Lyapunov exponent features of each point in the grid are represented by assigning different colors. In Figure~\ref{fig:1}, blue color represents parameter sets for which the attractors are chaotic (positive Lyapunov exponent), while grey-scale represents parameter sets for which the attractors are periodic (negative Lyapunov exponent). In black we represent points of the superstable lines corresponding to periodic attractors for which the largest non zero Lyapunov exponents is a minimun inside a periodic window.

We observe, in Figure~\ref{fig:1}, that the parameter sets for which the Duffing oscillator behaves periodically are in aggregated periodic windows (gray scale area). Those periodical structures, also known as shrimps, have been well described in literature \cite{1,2,3}. In Figure~\ref{fig:1}(b), the red circle and the black plus symbol indicate examples of parameter sets whose orbits are, respectively, periodic and chaotic; these two orbits will be further considered in Section~\ref{sec:perduff}.

\begin{figure}[!ht]
\centering
\includegraphics[width=\columnwidth]{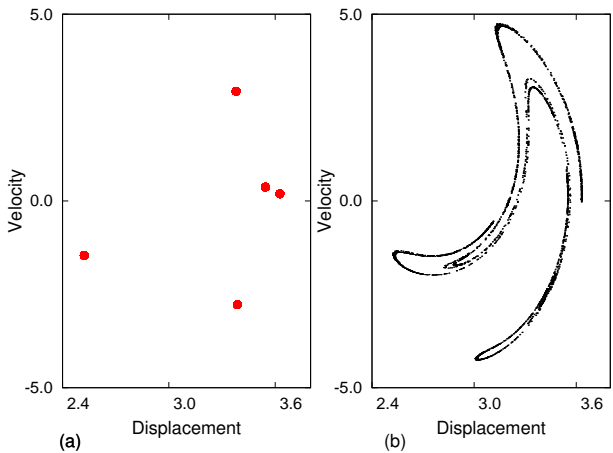}
\caption{ (Color online) Stroboscopic phase space of the Duffing oscillator. (a) Periodic orbit correspondent to the parameters $\beta=10.6780$ and $c=0.2670$. (red circle in Figure 1(b)). (b) Chaotic attractor correspondent to the parameters $\beta = 10.7020$ and $c = 0.2670$ (black plus symbol in Figure 1(b)).}
\label{fig:2}
\end{figure}

To show the different behavior of the two attractors associated with the two parameter sets marked in Figure~\ref{fig:1}(b), we obtain a stroboscopic map by collecting the velocity and the displacement at (Time-$2\pi / \omega$). In Figure~\ref{fig:2}(a), we show the periodic orbit correspondent to the parameters represented in red inside the periodic window amplified in Figure~\ref{fig:1}(b). Similarly, in Figure~\ref{fig:2}(b), we show the chaotic attractor for parameters represented by a plus symbol in Figure~\ref{fig:1}(b).

We call the attention, in Figure~\ref{fig:2}, that both periodic and chaotic attractors are in the same phase space region. Next, in Section~\ref{sec:perduff}, we analyze how these two orbits are modified by a weak periodic perturbation.

\section{Introducing a weak perturbation in the Duffing oscillator}
\label{sec:perduff}

We introduce a weak harmonic perturbation by adding a second harmonic term in the original Duffing driven. The perturbation amplitude is taken as a control parameter of the system. On the other hand, the perturbation frequency is settled in an integer ratio of the original system frequency. Similar procedures have been already reported in literature \cite{20,21,22}. So, the time evolution of this perturbed system is determined by solution of the following dimensionless equation:

\begin{equation}
\ddot{x}+c \dot{x}+x^3=\beta \cos (\omega t) + \alpha sin (\Omega t).
\label{eq:2}
\end{equation}

Here, $\alpha$ is the weak perturbation amplitude $(\alpha << \beta)$ and $\Omega$ is the perturbation frequency fixed at $\Omega=2\omega$. Other rational multiples of $\omega$ could be used to $\Omega$ \cite{17,21}. 

\begin{figure*}[!htp]
\centering
\includegraphics[width=17cm,height=9cm]{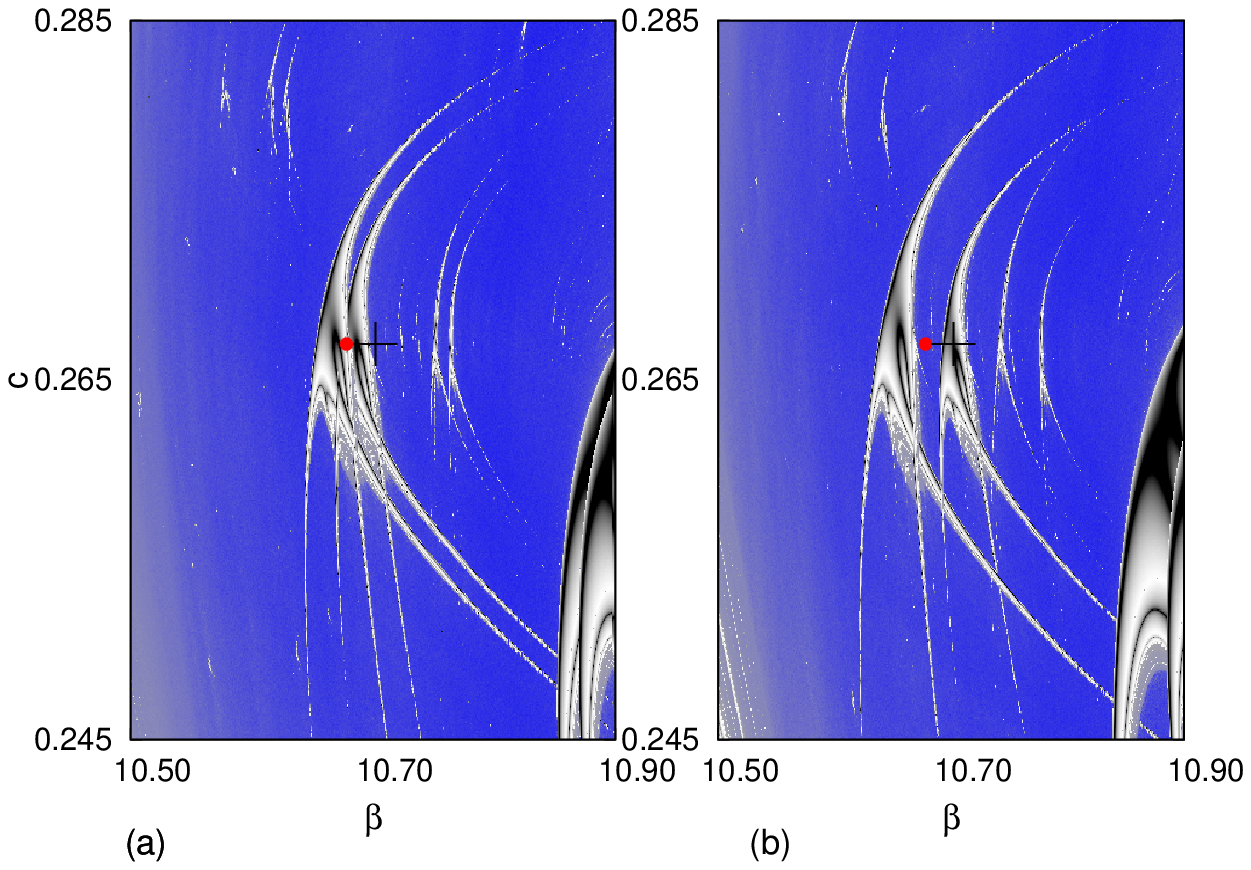}
\caption{ (Color online) Bi-dimensional ($c$ $\times$ $\beta$) parameter space of the perturbed Duffing oscillator for $\omega = 1.0$ and $\Omega = 2.0$. The perturbation amplitude is $\alpha = 0.04$ in (a) and $\alpha = 0.08$ in (b).}
\label{fig:3}
\end{figure*}

In Figure~\ref{fig:3}, we present the parameter space of the perturbed Duffing oscillator for two different perturbation amplitudes $\alpha = 0.04$ and $\alpha = 0.08$. By comparing the parameter spaces of Figure~\ref{fig:1}(b) and Figures~\ref{fig:3}(a,b), we note the existence of replicated periodic windows. In particular, the central periodic window of Figure~\ref{fig:1}(b) is duplicated in Figure~\ref{fig:3}. Moreover, the new periodic windows appear slightly displaced in Figures~\ref{fig:3}(a) and \ref{fig:3}(b). Thus, in Figure~\ref{fig:3}(b), for the perturbation amplitude $\alpha = 0.08$, the black plus marked point is inside the replicated periodic window. In other words, the chaotic attractor showed in Figure~\ref{fig:2}(a) has been changed to a periodic one.

\begin{figure}[!ht]
\includegraphics[width=\columnwidth]{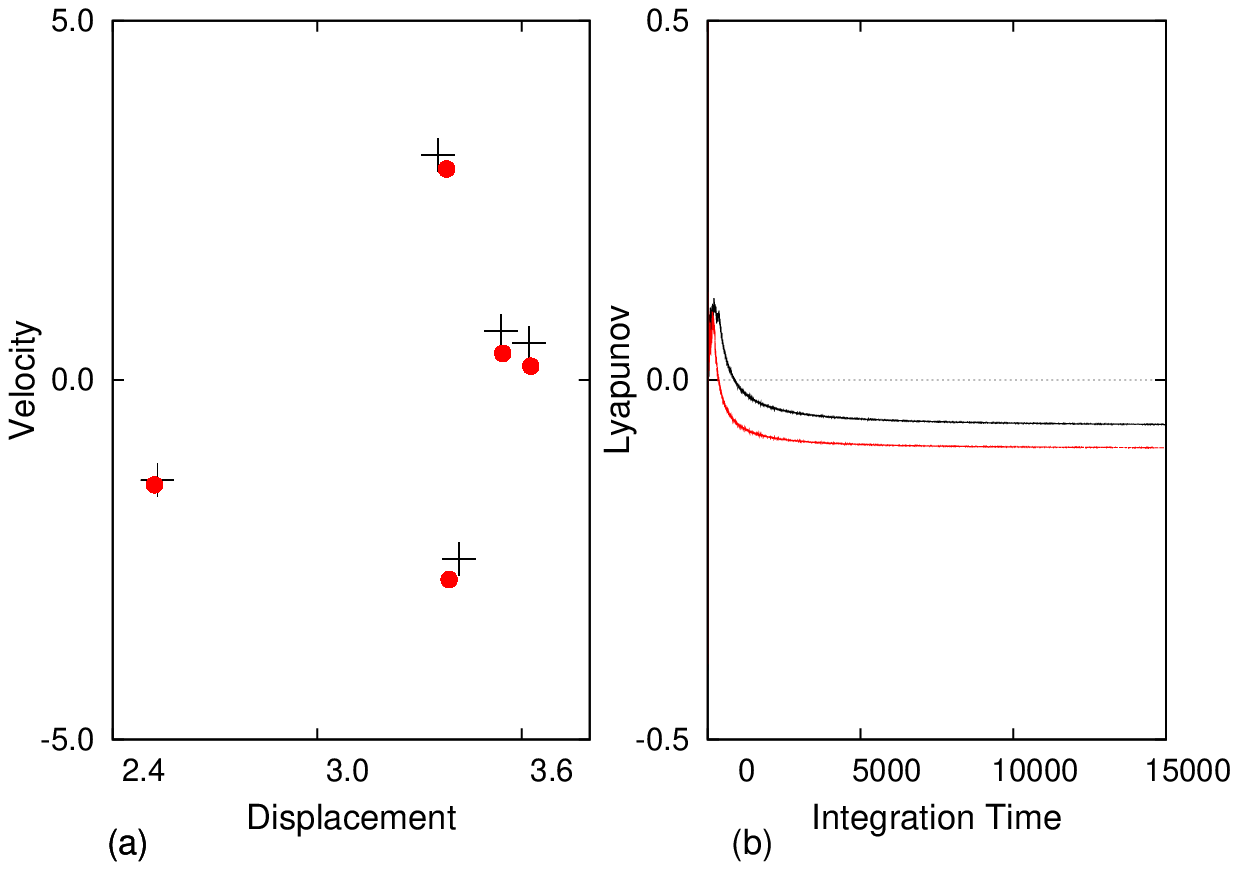}
\caption{ (Color online) (a) The red circle symbol denotes the periodic orbit for $\beta=10.6780$, $c=0.2670$ and $\alpha=0.0$, the same orbit of Figure 2(b). The black plus symbol denotes the controlled orbit for $\beta = 10.7020$, $c = 0.2670$, $\Omega=2.0$, and $\alpha=0.08$. (b) The red line indicates the largest Lyapunov exponent of the unperturbed periodic orbit. The black line indicates the largest Lyapunov exponent of the controlled orbit.}
\label{fig:4}
\end{figure}

Next, we discuss the alteration produced by the weak perturbation on the chaotic orbit shown in Figure~\ref{fig:2}(b). For this purpose, in Figure~\ref{fig:4}(a), we show the stroboscopic phase space of the new periodic orbit (in black plus symbols), that substitutes the former chaotic orbit shown in Figure~\ref{fig:2}(b). Additionally, to compare similarities between perturbed and unperturbed orbits, in Figure~\ref{fig:4}(a), we present (in red points) the unperturbed periodic orbit whose parameters are indicated in red in Figure~\ref{fig:1}(b). In Figure~\ref{fig:4}(b), we show the convergence of the largest non zero Lyapunov exponent of the two orbits of Figure~\ref{fig:4}(a). We recognize, in Figure~\ref{fig:4}(a), that the new perturbed periodic orbit and the previous periodic orbit have the same periodicity and similar pattern. Moreover, the largest non zero Lyapunov exponent of both orbits also have a similar convergence.

We investigate the relation between the new perturbed periodic orbit and a possible unstable periodic orbit embedded in the chaotic sea. Thus, for the parameters marked by a plus symbol in Figure~\ref{fig:1}(b), we integrate the unperturbed equation for a short integration time, starting with initial conditions settled in the unperturbed periodic orbit with parameters marked by red circle in Figure~\ref{fig:1}(b). 

\begin{figure}[!ht]
\includegraphics[width=\columnwidth]{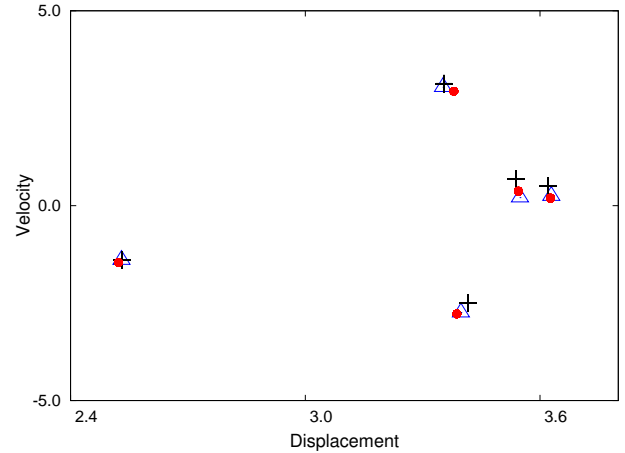}
\caption{ (Color online) The blue triangle indicates the chaotic orbit under a short integration time for $\beta = 10.7020$, $c = 0.2670$ and $\alpha=0.0$. The black plus symbol indicates the controlled periodic orbit for $\beta = 10.7020$, $c = 0.2670$, $\alpha = 0.08$ and $\Omega=2.0$. The red circle symbol indicates the unperturbed periodic orbit for $\beta=10.6780$, $c=0.2670$ and $\alpha=0.0$.}
\label{fig:5}
\end{figure}

In Figure~\ref{fig:5}, the blue triangle symbol denotes the chaotic orbit in the short integration time, the red circle symbol denotes unperturbed existent periodic orbit, and the black plus symbol denotes the perturbed periodic orbit. In Figure~\ref{fig:5}, we observe that for a short integration time, the chaotic orbit remains in the neighbor of the perturbed and the unperturbed periodic orbit.

The analysis of results shown in Figures~\ref{fig:4} and \ref{fig:5} indicate that the periodic window replication described in this work gives rise to new periodic windows whose parameters correspond to both stable and unstable periodic orbits with the same periodicity and pattern found in previous existing unperturbed oscillator.

\section{Weak periodic perturbation in the Josephson Junction oscillator}
\label{sec:pertjose}

The superconducting Josephson junction is usually modeled by an electronic circuit equation \cite{28}. Considerable efforts has been devoted to understand the onset of chaos in this system and its  critical parameters for this transition \cite{29,30,31}. The system is described by a pendulum-like dimensionless equation \cite{22}:

\begin{equation}
\ddot{\phi}+G \dot{\phi}+\sin \phi= I + A \sin (\omega t).
\label{eq:3}
\end{equation}

Here, the parameter $G$ gives the amplitude of the system damping, $I$ is the direct current component of the circuit, $\omega$ is the natural system frequency settled at $\omega=0.25$, while $A$ is the alternating current component of the circuit and the system forcing amplitude. 

\begin{figure*}[!htp]
\includegraphics[width=17cm,height=9cm]{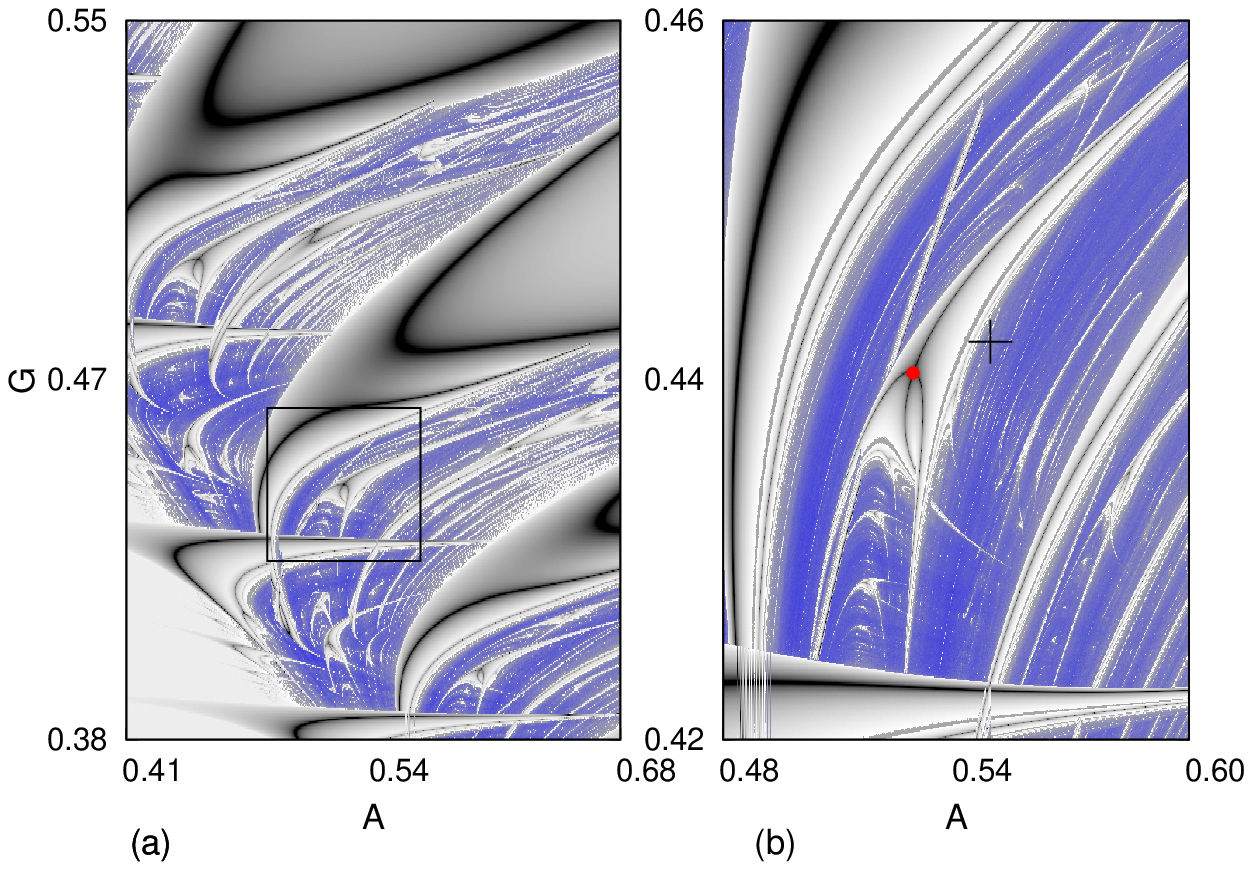}
\caption{ (Color online) (a) Bi-dimensional ($G$ $\times$ $A$) parameter space of the unperturbed Josephson oscillator, the natural system frequency is settled for $\omega = 0.25$, the direct current component is settled for $I=0.905$. (b)  Magnification of the squared area. The red circle indicates the parameters $A=0.5289$ and $G=0.4403$, the black plus symbol indicates the parameters $A=0.5488$ and $G=0.4416$.}
\label{fig:6}
\end{figure*}

In Figure~\ref{fig:6}, we obtain the bi-dimensional parameter space ($G$ $\times$ $A$) of the Josephson oscillator given by Equation~\ref{eq:3}. In this figure, blue points correspond to positive Lyapunov exponent (chaotic behavior), while grey- scale points correspond to negative Lyapunov exponent (periodic behavior). The black region corresponds to the superstable lines.

We verify, in Figure~\ref{fig:6}, the existence of aggregated periodic windows in the parameter space of the Josephson oscillator. As in Figure~\ref{fig:1}(b), the red circle and the black plus symbol indicate in Figure~\ref{fig:6}(b) examples of parameters for which the correspondent orbits are, respectively, periodic and chaotic.

\begin{figure}[!ht]
\includegraphics[width=\columnwidth]{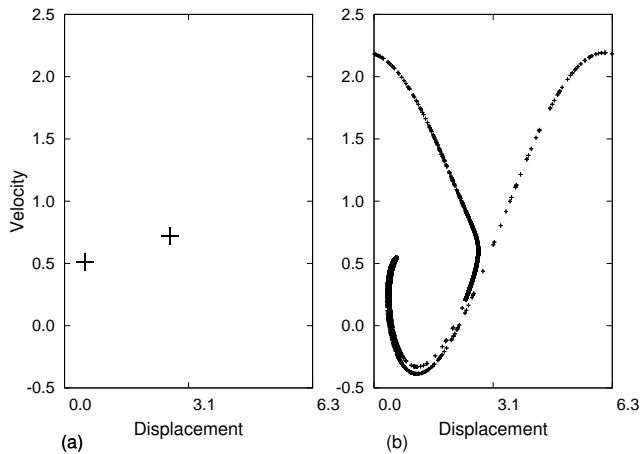}
\caption{Stroboscopic phase space of Josephson oscillator. (a) Periodic orbit correspondent to the parameters $A=0.5488$ and $G=0.4416$. (red circle in Figure 6(b)). (b) Chaotic attractor correspondent to the parameters $A=0.5488$ and $G=0.4416$ (black plus symbol in Figure 6(b)).}
\label{fig:7}
\end{figure}

In Figures~\ref{fig:7}(a) and \ref{fig:7}(b) we present, respectively, the periodic and the chaotic orbits for the two parameter sets indicated in Figure~\ref{fig:6}(b), to be further analyzed in this section.

For the Josephson oscillator, similarly to the perturbed Duffing oscillator (Considered in Section 3), a harmonic term is added to the original one:

\begin{equation}
\ddot{\phi}+G \dot{\phi}+\sin \phi= I + A \sin (\omega t) + \alpha \sin (\Omega t).
\label{eq:4}
\end{equation}

where, $\alpha$ is the perturbation amplitude and $\Omega$ is the perturbation frequency fixed at $\Omega=\omega/2$.

\begin{figure*}[!htp]
\includegraphics[width=17cm,height=9cm]{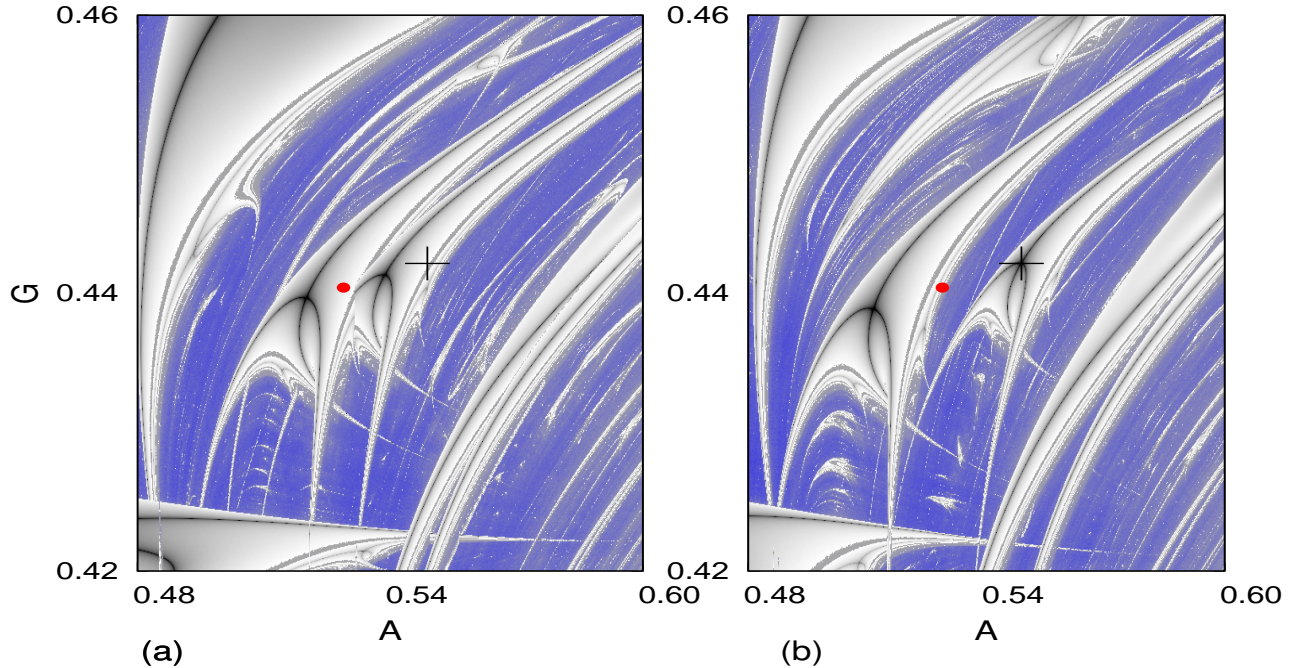}
\caption{ (Color online) Bi-dimensional ($G$ $\times$ $A$) parameter space of the perturbed Josephson oscillator, the natural system frequency and the perturbation frequency are settled, respectively, for $\omega = 0.25$ and $\Omega = 0.125$. (a) The perturbation amplitude is settled for $\alpha = 0.005$. (b) The perturbation amplitude is settled for $\alpha = 0.01$.}
\label{fig:8}
\end{figure*}

In Figure~\ref{fig:8} we show the parameter space of the perturbed Josephson equation. Similarly to what was noticed in Figure~\ref{fig:3} for the Duffing oscillator, in this case the existence of replicate periodic windows in the modified parameter space is also observed. Moreover, comparing Figures~\ref{fig:6} and \ref{fig:8}, we see that the parameter set indicated by a plus symbol in these two figures represent a chaotic attractor in Figure~\ref{fig:6}(b), for the unperturbed oscillator, and a periodic one, in Figure~\ref{fig:8}(b), for the perturbed oscillator.

\begin{figure}[!ht]
\includegraphics[width=\columnwidth]{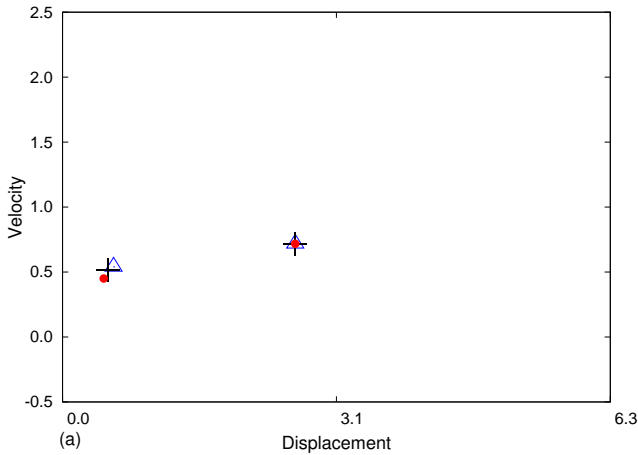}
\caption{ (Color online) The blue triangle indicates the chaotic orbit under a short integration time for $A=0.5488$, $G=0.4416$ and $\omega=0.25$. The black plus symbol indicates the controlled periodic orbit for $A=0.5488$, $G=0.4416$, $\omega=0.25$ and $\Omega=0.125$. The red circle symbol indicates the previous periodic orbit for $A=0.5488$, $G=0.4416$ and $\omega=0.25$.}
\label{fig:9}
\end{figure}

Next, to show how these orbits are modified by the considered perturbation with amplitude $\alpha = 0.01$, we show in Figure~\ref{fig:9} the stroboscopic phase space of the perturbed Josephson oscillator with three orbits. Thus, in Figure~\ref{fig:9}, the black plus symbol represents the controlled periodic orbit indicated by a black plus symbol in the parameter space shown in Figure~\ref{fig:8}(b), with the same parameters of the former chaotic orbit  marked with the black plus symbol in Figure~\ref{fig:6}(a). The blue triangles indicate the chaotic orbit (whose parameter set is indicated in Figure~\ref{fig:6}(b)) under a short integration time for the initial condition fixed on the periodic attractor. The red circle symbol denotes unperturbed periodic orbit.
 
Thus, the results shown in Figures \ref{fig:6} - \ref{fig:9} confirm the evidence, found for the Duffing oscillator, that the new periodic orbits are similar to previous stable and unstable periodic orbits of the unperturbed oscillator.

\section{Conclusions}
\label{concl}

We investigate the control of chaos for two driven oscillators by a weak harmonic forcing. To identify periodic and chaotic regions in the bi-dimensional parameter space, we compute the largest non zero Lyapunov exponents for the attractors in the considered parameter ranges. We identify shrimp-shaped periodic windows immersed into a chaotic region. The parameter space is much modified whenever a weak amplitude forcing is applied. New similar periodic windows arise in the neighborhood of the original windows. We verify that periodic orbits are similar (and with the same periodicity) for parameters inside the original and the new periodic windows. These results are similar to those obtained for the driven impact oscillator \cite{17}.

By analyzing stroboscopic maps of unperturbed and perturbed attractors we find evidences that the new reported periodic windows are formed by parameters for which the observed new periodic attractors are similar to preexisting stable and unstable periodic orbits. Therefore, we conjecture that the replicate periodic windows reported in this work are associated with chaos control and reproduce further other periodic orbits of the considered oscillators.

\section{Acknowledgements}

 	This work was made possible by partial financial support from the following Brazilian government agencies: FAPESP, CNPq, and Capes.

\end{document}